\documentclass[aps,prl,twocolumn,superscriptaddress]{revtex4}
\usepackage{amsfonts}
\usepackage{amsmath}
\usepackage{amssymb}
\usepackage{graphicx}
\usepackage{color}
\usepackage{float}

\begin{document}
\title{ Comparison of  time reversal symmetric  and asymmetric nano-swimmers oriented with an electric field in soft matter}


\author{G. Rajonson}
\affiliation{ Laboratoire de Photonique d'Angers EA 4464, Universit\' e d'Angers, Physics Department,  2 Bd Lavoisier, 49045 Angers, France.\\
*Electronic mail: victor.teboul@univ-angers.fr}

\author{D. Poulet}
\affiliation{ Laboratoire de Photonique d'Angers EA 4464, Universit\' e d'Angers, Physics Department,  2 Bd Lavoisier, 49045 Angers, France.\\
*Electronic mail: victor.teboul@univ-angers.fr}

\author{M. Bruneau}
\affiliation{ Laboratoire de Photonique d'Angers EA 4464, Universit\' e d'Angers, Physics Department,  2 Bd Lavoisier, 49045 Angers, France.\\
*Electronic mail: victor.teboul@univ-angers.fr}

\author{V. Teboul}
\email{victor.teboul@univ-angers.fr}
\affiliation{ Laboratoire de Photonique d'Angers EA 4464, Universit\' e d'Angers, Physics Department,  2 Bd Lavoisier, 49045 Angers, France.\\
*Electronic mail: victor.teboul@univ-angers.fr}


\begin{abstract}

Using molecular dynamics simulations we compare the motion of a nano-swimmer based on Purcell's suggested motor with a time asymmetrical cycle with the motion of the same molecular motor with a time symmetrical cycle. We show that Purcell's theorem still holds at the nanoscale, despite the local structure and the medium's fluctuations. 
Then, with the purpose of both orienting the swimmer's displacement and increasing the breakdown of the theorem, we  study the effect of an electric field on a polarized version of these swimmers.
For small and large fields, the time asymmetrical swimmer is more efficient, as suggested by Purcell.
However we find a field range for which Purcell's theorem is broken for the time symmetric motor.  Results suggest that the  {\color{black} breakdown of the theorem}  is arising from the competition of the orientation field and Brownian forces, while for larger fields the field destroys the effect of fluctuations restoring the theorem. 

\end{abstract}

\maketitle
\section{ Introduction}

Since the emergence of nanotechnology, the design and properties of synthetic molecular motors and machines are the subjects of active research\cite{moto1,moto2,moto3,moto4,moto5,moto6,moto7,motor0,motor1,motor2,motor3,motor4,motor5,motor6,motor7,motor8,motor10,motor11,motor12,motor13,motor14,motor15,motor16,motor17,pccp,motor18}. 
Molecular motors applications range from medicine to engineering, sensing, the control of transport mechanisms, crystallization, hydrophobicity, optical properties, and actuation to name a few.
Among molecular motors, nano-swimmers are of particular interest for medical applications. 
Biological molecular motors are quite complex molecular systems. 
Synthetic molecular motors that have been created up to date are also still  relatively complex.  
Creating simpler motors is however of large interest as it will permit to produce them more easily in large number, and can lead to more stable motors.

Due to Brownian motion and low Reynolds number, creating nano-swimming molecular motors is quite a challenge.
Brownian motion hinders directional motor's motion by incessant interactions with the medium,  and induces fluctuations in the environment local structure and dynamics.
Moreover, because the Reynolds number $Re$ is proportional to the characteristic length-scale of the system, at the nanoscale $Re$ is small.
As $Re$ measures the ratio of kinetic to viscous forces, at the nanoscale viscous forces govern the dynamics, hindering the motor's displacements.
Eventually, low Reynolds number lead to Purcell's theorem\cite{scallop1,purc} that forbids the motion of time symmetrical nano-swimmers.

While Brownian motion introduces some  {\color{black} breakdown \cite{purc}} of the theorem due to fluctuations in the environment, previous studies found nevertheless that the theorem mostly holds at the nanoscale\cite{prefold,scallop13b}. 
Any attempt to create a simple mono-molecular motor  will  have to  {\color{black} breakdown \cite{purc}} the Purcell theorem to be efficient.
Various breakdowns of the theorem were reported \cite{scallop2,scallop4,scallop5,scallop6,scallop7,scallop8,scallop9,scallop11,scallop12,scallop13,scallop13b,scallop13c,scallop14,scallop15,prefold}.
 {\color{black} A few of these breakdowns appear for rapid flapping motors\cite{scallop6,scallop4,scallop14,scallop15} while most others are due to specific environments.}
{\color{black}  Diffusion of the motor can also occur due to fluidization of the medium by the motor's flaps}, the medium diffusion then carrying the motor\cite{md16,scallop3,cage,rate,carry,scallop10}. This effect called photofluidization \cite{flu1,flu2,flu3,flu4} for azobenzene motor molecules is of {\color{black} particular interest due to its relation with the long standing problem of the glass-transition \cite{gt1,gt2}.}

Purcell suggests in his paper\cite{scallop1} more complicated mechanisms to bypass the theorem. These mechanisms need at least two hinges, with foldings that have to be non-symmetrical in a time reversal transformation. 
In this paper we test the effect of time symmetry, comparing two nano-swimmers based on Purcell's suggested motor.
 One of our swimmers uses the time asymmetrical succession of flaps proposed by Purcell to  break the theorem, and the other one uses time symmetrical succession of flaps\cite{swim}.

\begin{figure}
\centering
\includegraphics[height=4.8 cm]{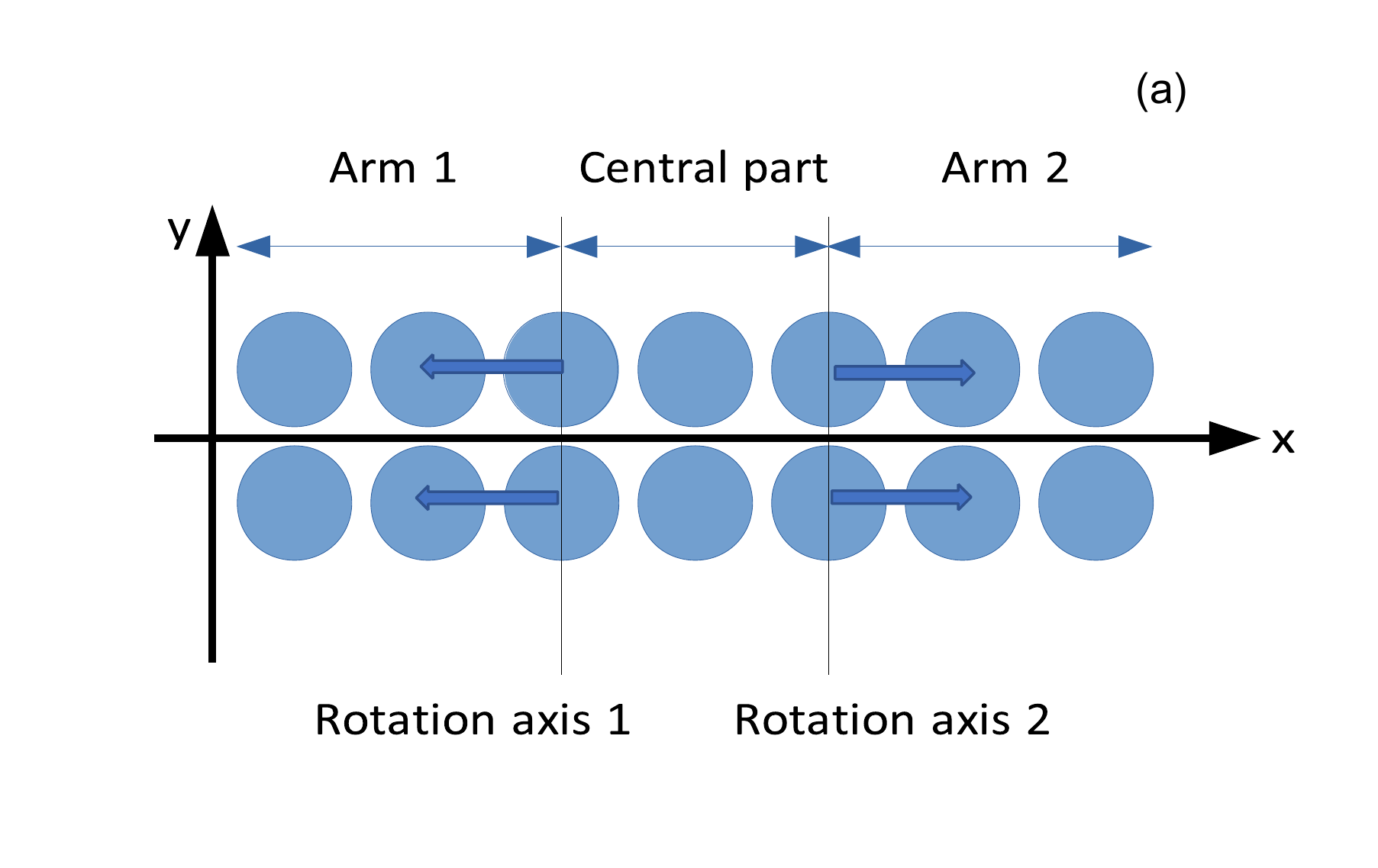}
\includegraphics[height=5.8 cm]{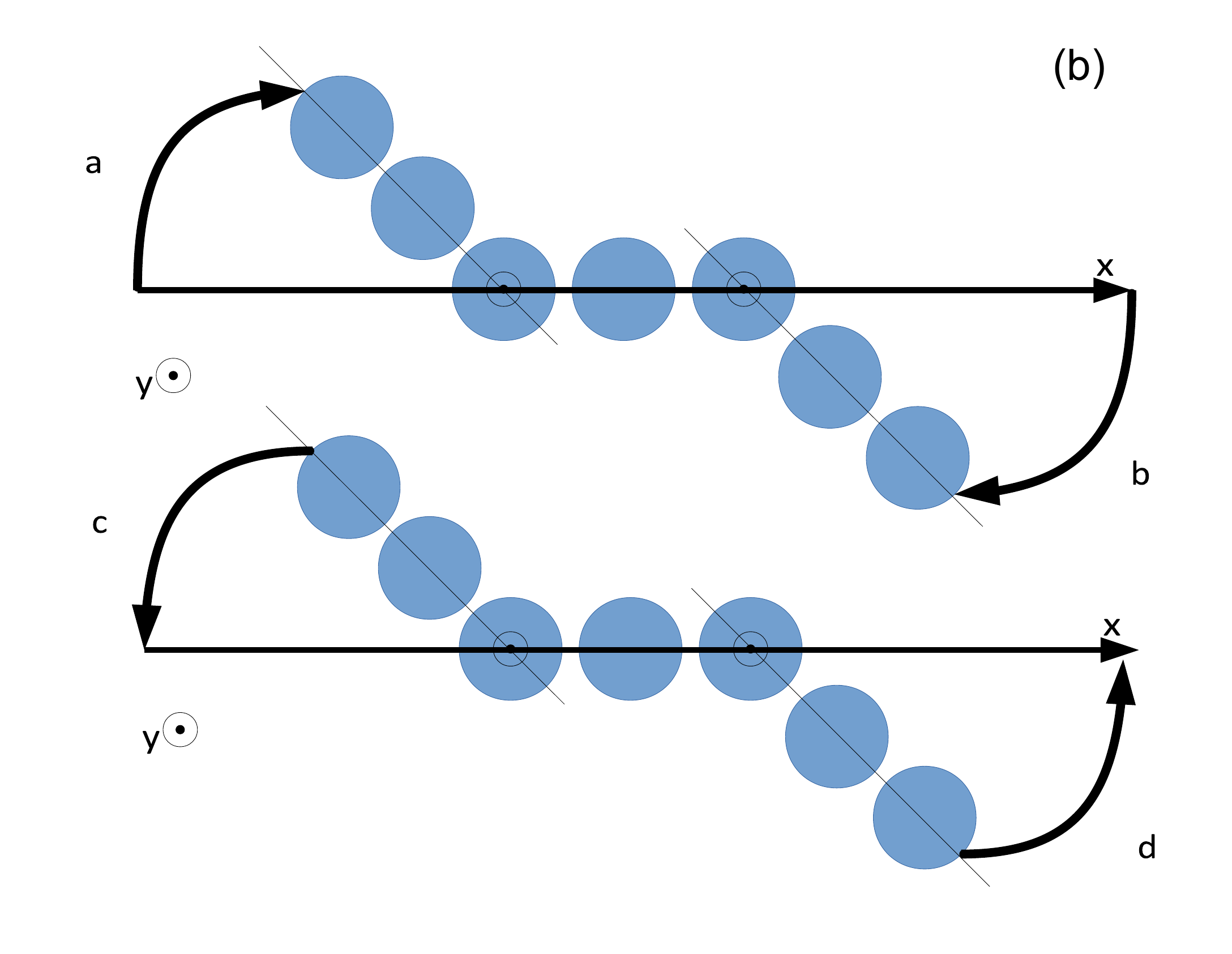}
\caption{(color online)  {\color{black} (a) Picture of the flat nano-swimmer before any arm's motion. Local coordinates $x$ and $y$ define the plane of the motor. The external forces are applied on the $4$ atoms of the rotation axis in the $-X$  and $X$ directions of the laboratory coordinates, so that the sum of the forces on the motor is null. For large fields, the motor eventually orients itself on the direction of the field and the $x$ motor's axis is then superimposed on the $X$ laboratory axis. The arrows indicate the external force field in that configuration.
Note that when the force field is applied, the motor's reversed orientation in the $-X$ direction is unstable. 
(b) The $4$ steps, beginning from an initially flat swimmer are described in the Figure. The order of the flaps is (abcd) for the time asymmetric nano-swimmer and (abdc) for the time symmetric one.
{\color{black} The time asymmetric swimmer is  based on the motor suggested by Purcell to break his theorem} due to time asymmetry while the other one is the same motor but with time symmetry; that is reversing time leads to the same succession of steps for the time symmetric motor but not for the time asymmetric one.
 The swimmers are adapted from Purcell's motor,  so that they can be experimentally engineered from two attached photo-isomerizing azobenzene molecules.
As explained by Purcell\cite{scallop1} the motor main direction of motion is the $x$ local axis.}
} 
\label{f00}
\end{figure}

\begin{figure}
\centering

\includegraphics[height=5.8 cm]{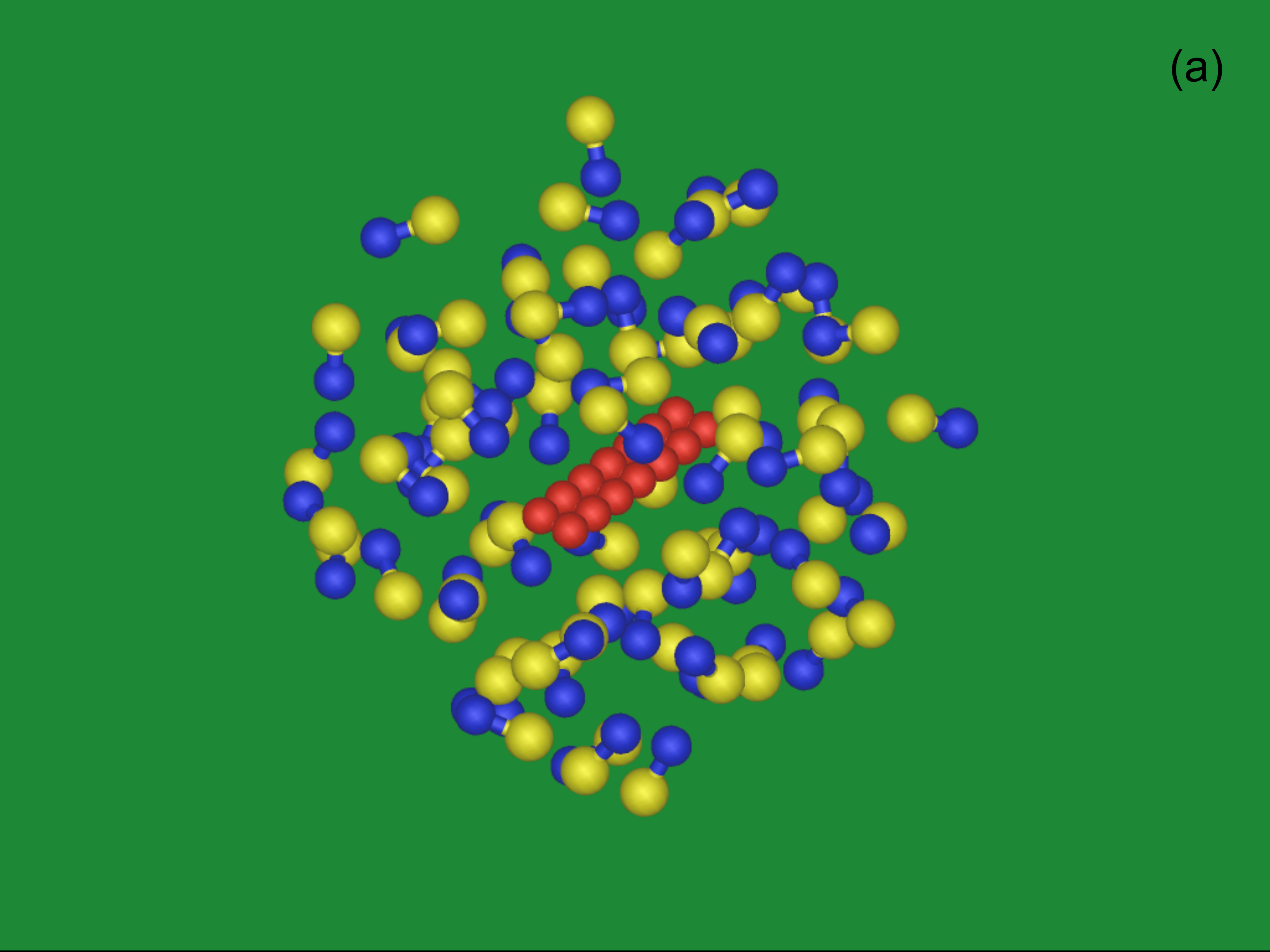}
\includegraphics[height=5.8 cm]{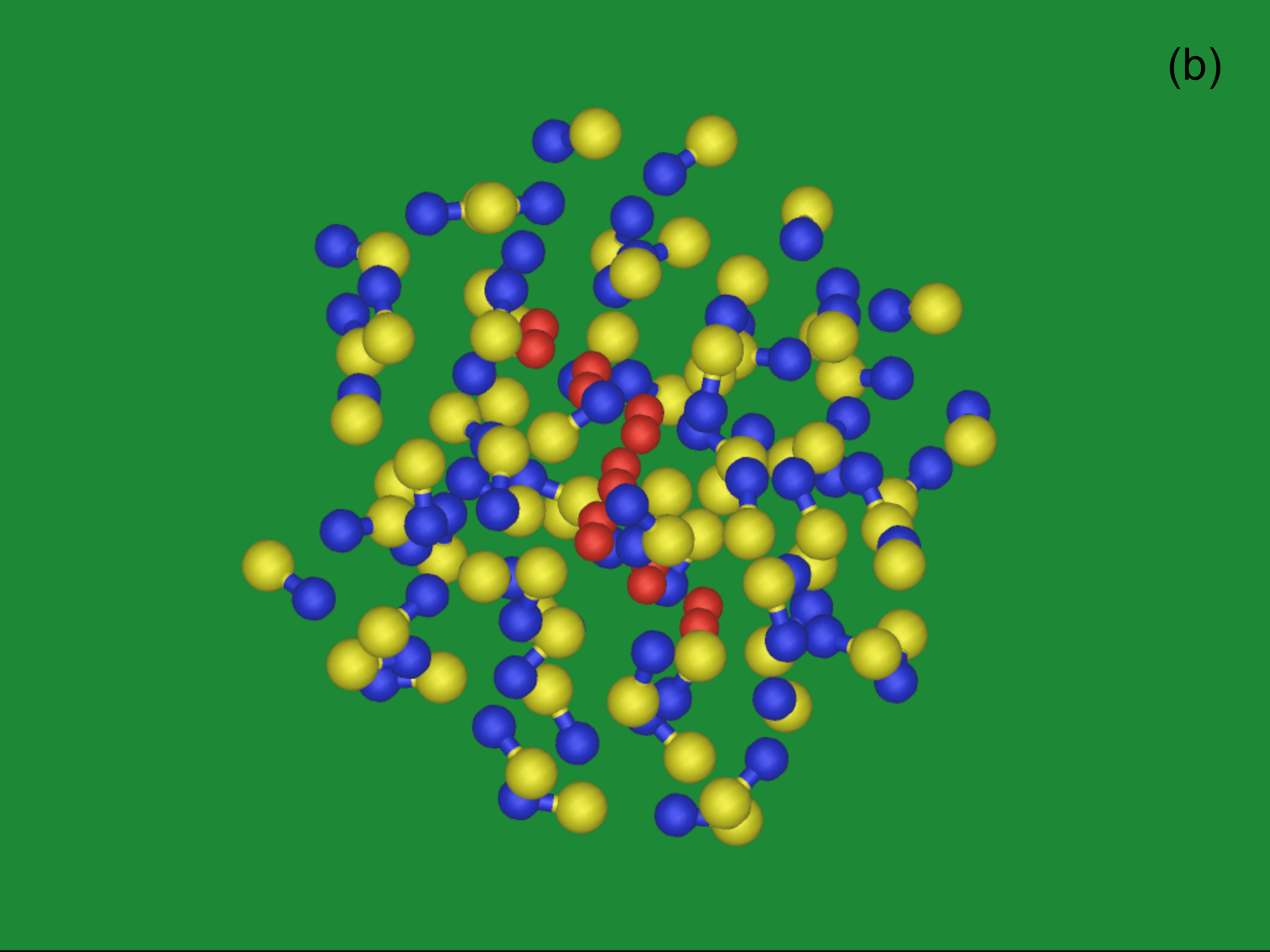}

\caption{(color online)  
{\color{black}
Snapshots of the motor and molecules surrounding it at a distance $r<10$\AA, with arbitrary colors. 
(a) The motor  is in its flat configuration.
(b) The two motor's arms are folded.
} }
\label{f01}
\end{figure}

We describe our two nano-swimming motors in Figures \ref{f00} and \ref{f01}.
 Our swimmers are adapted from Purcell's motor,  so that they can be experimentally engineered from two attached azobenzene molecules, that do have the property of photo-izomerization\cite{flu5,azo1,azo2,azo2b,azo3,azo4,azo5,azo6,azo7,azo8,azo9}.
They are constituted of three parts, with the parts at the two extremities moving periodically in four steps.
The only difference between the two motors is the order of the successive motions of the flapping parts.
In contrast to the time asymmetric motor, for the time symmetric motor the successive motions are the same if time is reversed.
As suggested by Purcell,  the asymmetrical motor displacement is much larger due to the theorem.

Molecular dynamics and Monte Carlo simulations\cite{md1,md2,md2b,md4} are invaluable tools, together with model systems\cite{ms1,ms2,ms3,ms4,ms5}, theoretical calculations and experimental data to increase our understanding of unsolved problems in condensed matter physics\cite{keys,md3,md4b,md6,md7,md8,md9,md10,yld,md11,md12,md13,md14,md15,md16,c2}.
We  study in this work the possible breakdown of the theorem on the time symmetrical motor, with molecular dynamics simulations, using an external electric field on a polarized motor.
The purpose of the electric field  is to both breakdown the theorem and induce a preferential direction in the motor's displacement.
We then compare the effect of various electric fields on our two motors to understand the effect of time symmetry on the displacements.
Our medium is a model system based on Lennard-Jones potentials so that most quantities (temperature, molecules sizes, etc.) can be tuned to approximately represent a number of viscous media, with our results scaled by the corresponding factors.

\section{Calculation}


The reader will find details on our simulation procedure in  previous papers\cite{pccp,prefold,ariane}, however for convenience we will resume the simulation procedure.
 We use the Gear algorithm with the quaternion method\cite{md1} to solve the equations of motions with a time step $\Delta t=10^{-15} s$. Due to the release of energy from the motor, simulations where the motor is active are out of equilibrium.
We evacuate the energy created by the motor's folding, from the system with a Berendsen thermostat\cite{berendsen}. 
We use the NVT canonic thermodynamic ensemble as approximated by that simple thermostat   (see ref.\cite{finite2} for an evaluation of the effect of the thermostat on our calculations). 
{\color{black}Our simulations use  one motor molecule imbedded inside a medium constituted of $500$ linear molecules, in a cubic box  $30.32$ \AA\ wide.} We use periodic boundary conditions. 
 The molecules of the medium (host)\cite{ariane} are constituted of two rigidly bonded atoms ($i=1, 2$) at the fixed interatomic distance  {\color{black} $l_{h}$}$=1.73 $\AA$ $. These atoms interact with atoms of other molecules with the following Lennard-Jones potentials:

\begin{equation}
V_{ij}=4\epsilon_{ij}((\sigma_{ij}/r)^{12} -(\sigma_{ij}/r)^{6})   \label{e1}
\end{equation}

with the parameters\cite{ariane}: $\epsilon_{11}= \epsilon_{12}=0.5 KJ/mol$, $\epsilon_{22}= 0.4 KJ/mol$,  $\sigma_{11}= \sigma_{12}=3.45$\AA, $\sigma_{22}=3.28$\AA.
The mass of the motor is $M=420 g/mole$ (constituted of $14$ atoms, each one of mass $30g/mole$) and the mass of the host molecule is $m=1000 g/mole$ ($2$ atoms with a mass of $500g/mole$ each).
We model the motor with $14$ atoms in a rectangular shape constituted of two rows of  $7$ rigidly bonded atoms. 
The width of the swimmer is $L_{s}=4.4$\AA\ and its length $l_{s}=15.4$\AA. It is constituted of two flapping {\color{black} rigid} parts of length $l_{a_{0}}=l_{a_{1}}=5.7$\AA\ and a central {\color{black} rigid} non mobile part of length  $d=4$\AA\ on which we apply the opposite force fields polarized in the $X$ direction of sum null.  {\color{black}Notice that the flapping parts and the central part are three rigid bodies in the simulations.}
The length of the host molecule is $l_{h}=5.09$\AA\ and its width $L_{h}=3.37$\AA.
The motor's atoms interact with the medium's atoms using mixing rules and a Lennard-Jones interatomic potential on each atom of the motor, defined by the parameters: 
$\epsilon_{33}= 1.355 KJ/mol$,  $\sigma_{33}=3.405$\AA. 
We use the following mixing rules \cite{mix1,mix2}: 

\begin{equation}
\epsilon_{ij}=(\epsilon_{ii} . \epsilon_{jj})^{0.5}  ; 									
\sigma_{ij}=(\sigma_{ii} . \sigma_{jj})^{0.5}   \label{e2}
\end{equation}

  for the interactions between the motor and the host atoms. 
{\color{black}The medium (host) is a fragile liquid\cite{fragile1,fragile2}  that falls out of equilibrium in our simulations below $T=38 K$, i.e. $T=38$ K is the smallest temperature for which we can equilibrate the system when the motor is not active. 
As a result above that temperature the medium behaves as a viscous supercooled liquid in our simulations and below that temperature it behaves as a solid (as $t_{simulation}<\tau_{\alpha}$). 
The simulations of this work correspond to $T=30 K$, therefore a temperature for which the medium behaves as a solid when the motor is not active. 
Notice however that when active, the motor's motions induce a fluidization of the medium around it. 
We evaluate the glass transition temperature $T_{g}$ to be slightly smaller  $T_{g} \approx 28 K$, from the change of the slope of the potential energy as a function of the temperature.}
However as they are modeled with Lennard-Jones atoms, the host and motor potentials are quite versatile.
Due to that property, a shift in the parameters $\epsilon$ will shift all the temperatures by the same amount, including the glass-transition temperature and the melting temperature of the material.

The two motors and their different folding steps are described in Figures \ref{f00} and \ref{f01}.
Each folding is modeled as continuous, using a constant quaternion variation, with a folding time $\tau_{f}=0.5 ps$.
The total cycle period is constant and fixed for both motors at $\tau_{p}=600 ps$.
Each of the $4$ steps described in the Figure \ref{f01} has the same duration equal to $\tau_{p}/4=150 ps$ including the $0.5 ps$ folding or unfolding.

Our simulations are out of equilibrium, as the motor releases periodically some energy into the medium surrounding it.
That released energy is then extracted from our system by the  Berendsen thermostat, and thus doesn't increase the mean temperature of our medium.
However our system, while out of equilibrium, is in a steady state and is not aging. That behavior is obtained because the energy released by the motor into the medium is small enough and the time lapse between two stimuli large enough for the system to relax before a new stimuli appears.
In other words we are in the linear response regime\cite{pccp}.

Through this work we use the mean square displacements of the motor to measure its ability to diffuses.
The mean square displacement is defined as\cite{md1}:\\

\begin{equation}
\displaystyle{<r^{2}(t)>= {1\over N.N_{t_{0}}} \sum_{i,t_{0}} \mid{{\bf r}_{i}(t+t_{0})-{\bf r}_{i}(t_{0})} }\mid^{2}   \label{e130}   
\end{equation}

From the time evolution of the mean square displacement we then calculate the diffusion coefficient $D$  for diffusive displacements using the Stokes-Einstein equation:
\begin{equation}
\displaystyle{\lim_{t \to \infty}<r^{2}(t)>=6 D t}
\end{equation}

{\color{black} Notice that the external field induces a preferential direction of motion for the motor leading to super-diffusive displacements. 
Therefore the displacements have a diffusive and a super-diffusive component.
However, due to symmetry for the time symmetric motor we expect the super-diffusive behaviors to disappear for larger time scales.
To evaluate diffusion coefficients when super-diffusive behaviors are present, we use the diffusive (intermediate times) part of the mean square displacement  instead of the long time limit.
To be complete, we also calculate in our study  the super-diffusive interpolation of the motor's motion.



}

To quantify the breakdown of Purcell's scallop theorem, following ref.\cite{prefold} we define  a coefficient $\epsilon$ that we call the efficiency  of the motor's motion:

\begin{equation}
\displaystyle{\epsilon= {<r^{2}(n\tau_{p})>\over 2n<r^{2}(\tau_{p}/2)>} }    \label{e13bb}   
\end{equation}

where $\tau_{p}$ is the period,  $\tau_{p}/2$  the time lapse for only $2$  steps as described in Figure \ref{f01}, and $n$ the number of periods considered.
In this paper we have used $n=3$.
{\color{black} The motion's efficiency $\epsilon$ compares the motor's mean square displacement (MSD) after a number $n$ of periods (i.e. after the theorem has applied) to  the motor's  MSD  after half a period (i.e. before the theorem can apply).
In other words $\epsilon$ is the ratio of the observed displacement of the motor, to the displacement expected without the Purcell theorem effect.
} With $\epsilon$ definition, if the scallop theorem holds the efficiency $\epsilon=0$, while with a random motion we will obtain $\epsilon=1$, and a larger value of $\epsilon$ will mean that the motion is not random but has a  preferential direction.
Unless otherwise mentioned, the results displayed in this paper correspond to a temperature $T=30$ K in our model, a temperature for which the medium would be solid without the motor's stimuli.
{\color{black} However the motor's stimuli induce a fluidization of the medium around it. The fluidization mechanism\cite{flu1,flu2,flu3,flu4,md16} is of particular importance due to its relation with the glass-transition problem\cite{md16}. The effect of this fluidization on the motor's motion has been studied in previous papers \cite{scallop13b,scallop13c}.  
A motor carried by the medium due to fluidization will have an efficiency $\epsilon=1$ as it will not be subject to the Purcell theorem.
From the ratio of the motor's and medium's thermal diffusion coefficients at various temperatures, we estimate the part of the motor's motion due to the fluidization to be  $\leq 20$ percent, leading to a minimum value for the efficiency different from zero. }

\begin{figure}
\centering
\includegraphics[height=6.6 cm]{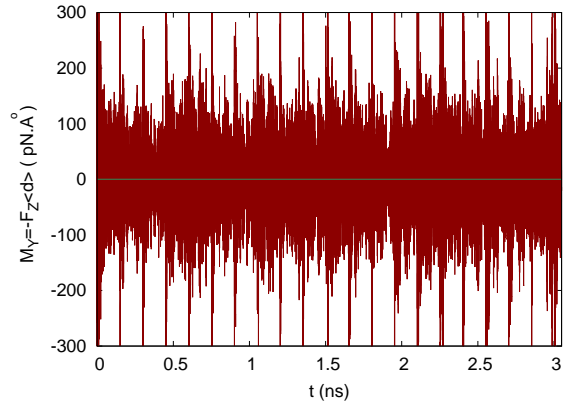}
\caption{{\color{black}(color online)  Time evolution of the $y$ component in the motor reference frame (see Figure \ref{f00}) of the force moment $M_{y}= - F_{z}$ $<d>$ for a time symmetric motor without external force field. The Brownian force moment is approximately equal to $100 pN.$\AA\ in the Figure, with some larger fluctuations. The periodic large peaks correspond to the force moments induced by the motor's flaps.\\} }
\label{ff00}
\end{figure}

{\color{black} The motor is subject to an external force field, leading to two opposite forces that apply in our calculations on the edges of the  central part of the motor, apart from a distance $d$ (see Figure \ref{f00}).
The physical parameter of importance in our system is thus the force moment (or torque) ${\bf M}={\bf dxF}$ applied on the motor.
However as this moment evolves with the motor's orientation, in our Figures we use the maximum moment ($M_{max}=\mid F \mid d$) as parameter.
Notice that the applied forces are equivalent experimentally to the effect of an external electric (or magnetic) field to a motor with a permanent dipolar electric (or magnetic) moment.}
We show a typical motor's moment time evolution in Figure \ref{ff00}. The larger peaks correspond to the flaps of the motor, while the Brownian noise  stays mainly around $100 pN$.\AA, with some larger fluctuations.

\section{Results and discussion}

\subsection{Does Purcell's theorem still hold at the nanoscale ?}

Purcell's theorem was established for a continuous medium, with Navier-Stokes hydrodynamic equations.
However at the nanoscale the medium is no longer continuous, because the medium's molecules are no longer small in comparison to the motor.
Also, at that length scale fluctuations in the environment are important, arising both from Brownian noise and from the motor's flapping arms perturbations.
Eventually, for amorphous soft matter and supercooled liquids, cooperative fluctuations (called dynamic heterogeneities) arise, increasing at low temperature.
For all these reasons, Purcell's theorem should be intrinsically broken at the nanoscale.
We will however see now that the theorem still holds at the nanoscale.

\begin{figure}
\centering
\includegraphics[height=6.6 cm]{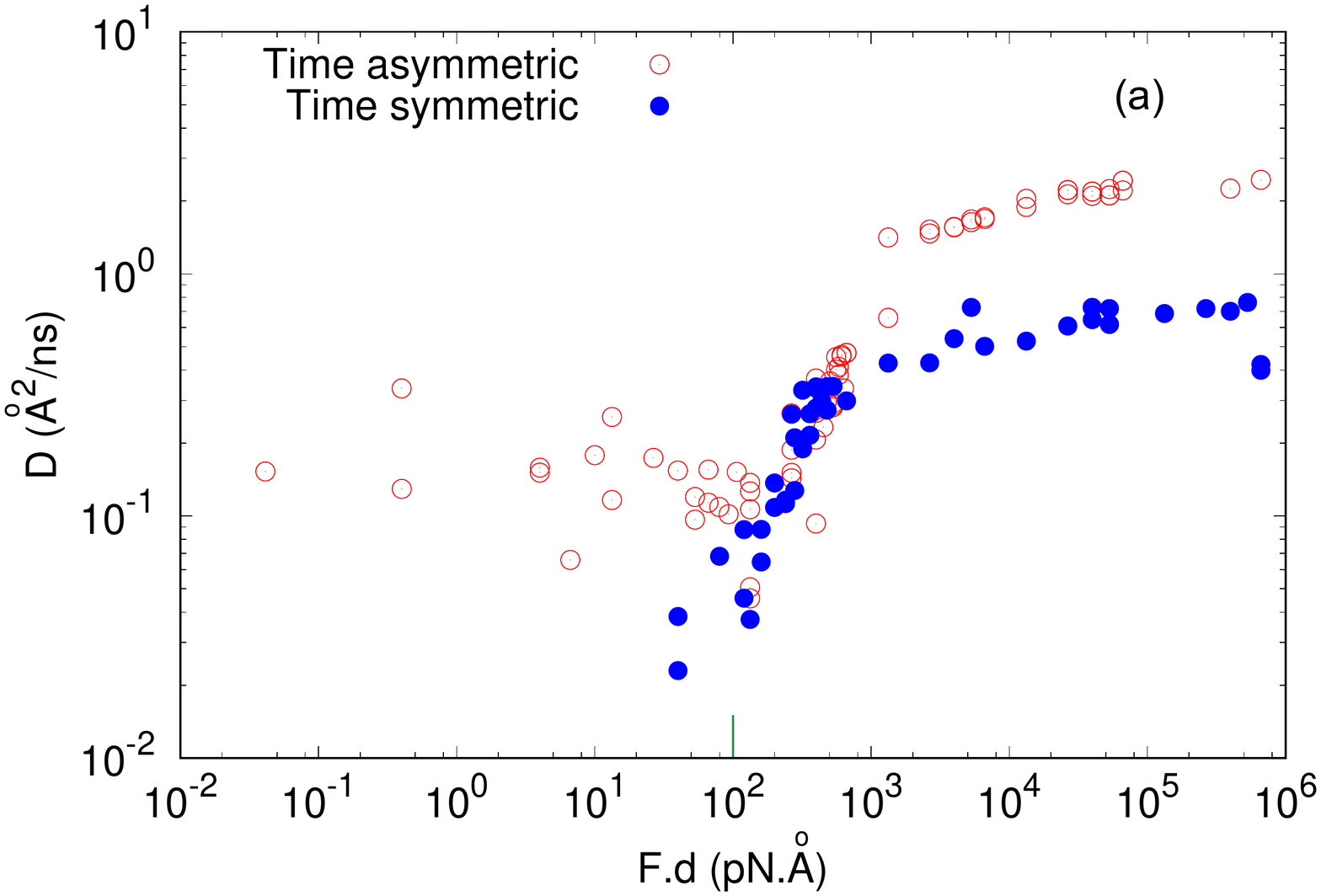}
\includegraphics[height=6.6 cm]{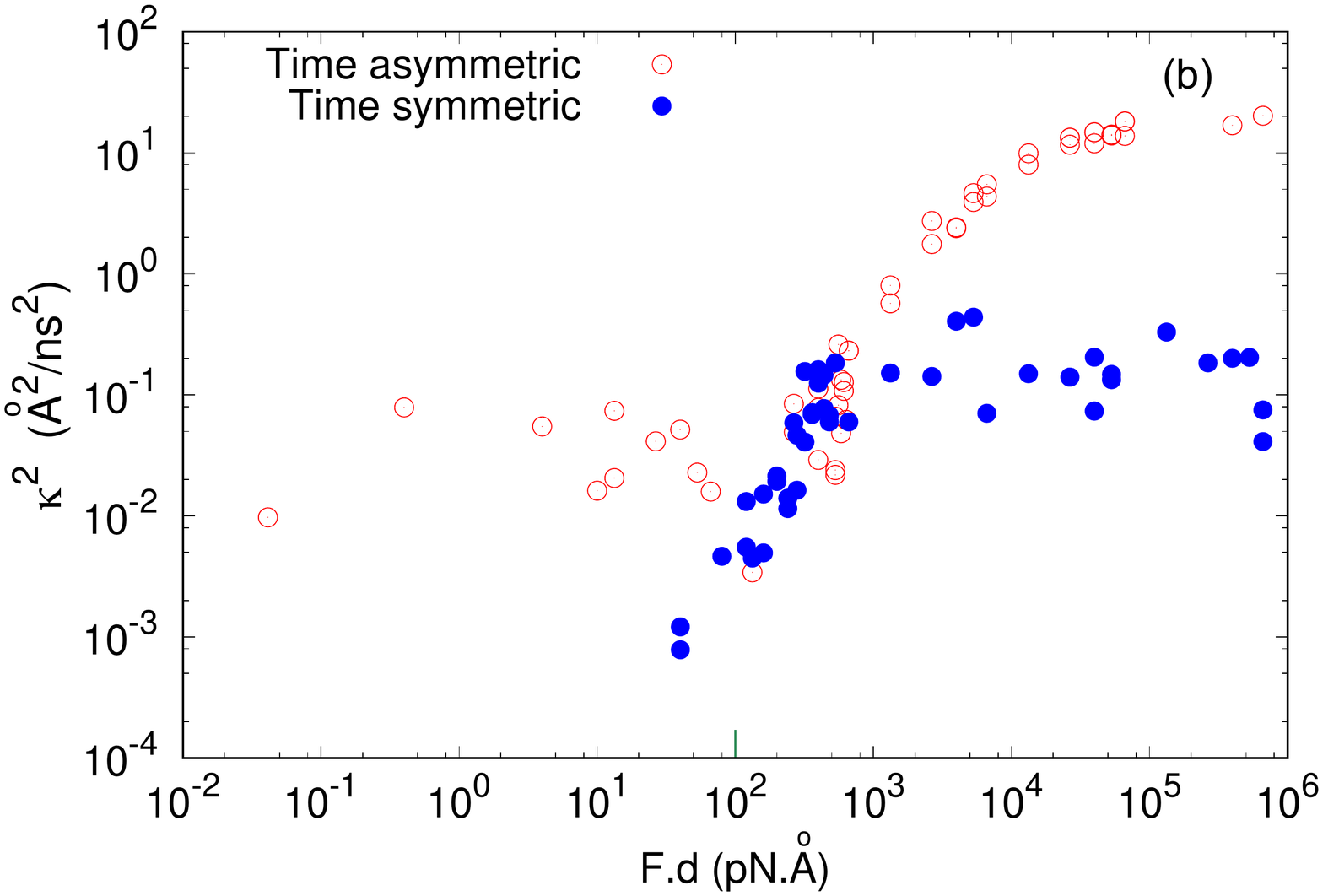}

\caption{(color online) (a) Diffusion coefficient $D$ versus the maximum moment $F.d$ for the time symmetric and time asymmetric motors.
The diffusion coefficients displayed here, are obtained from fits of the mean square displacement in the time range for which it is approximately linear. {\color{black}The small line shows the value of the Brownian noise (see Figure \ref{ff00}).
{\color{black} (b) Interpolation of the square of the mean motor's velocity $\kappa$ for super-diffusive displacements.}\\} }
\label{f3}
\end{figure}

We will now use the two motors diffusion coefficients displayed in Figure \ref{f3} to measure the breakdown of the scallop theorem.
The asymmetric motor is not constrained by the theorem, while the symmetric motor is.
Therefore a comparison between the diffusion behavior of the two motors shows the domain of  departure from  the theorem by the symmetric motor.
Note also that the presence of diffusion is in itself a  proof of breakdown of the theorem.

Figure \ref{f3}a shows the diffusion coefficient  {\color{black} $D$ and Figure \ref{f3}b the mean velocity $\kappa$ (for super-diffusive motions)} of the two motors as a function of the external force field torque.
There is a threshold on the external force below which the time symmetric (i.e. reversible in time) motor doesn't move while the time asymmetric motor does.
As the only difference between the two motors is the intrinsic  breakdown of Purcell's theorem by the asymmetric motor, these results show that Purcell's theorem still holds at the nanoscale in our conditions of study.

\subsection{Field-induced breakdown of Purcell's theorem}

\begin{figure}
\centering
\includegraphics[height=6.6 cm]{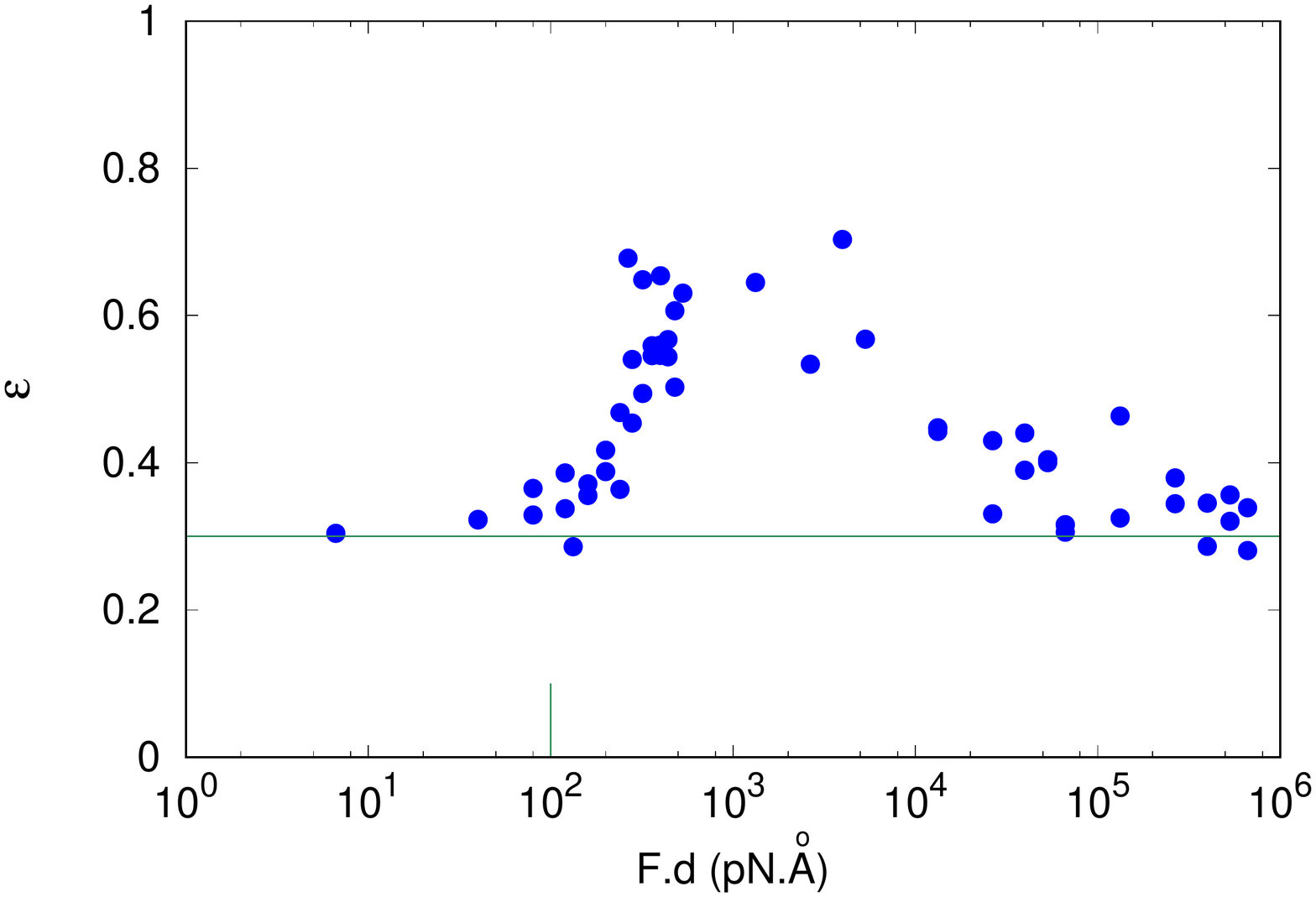}
\caption{(color online) Mobility efficiency coefficient $\displaystyle{\epsilon= {<r^{2}(n\tau_{p})>\over 2n<r^{2}(\tau_{p}/2)>} }$ with $n=3$, versus the maximum moment for the time symmetric motor. {\color{black}The small line at $100 pN.$\AA\ shows the value of the Brownian noise (see Figure \ref{ff00}).}} 
\label{f5}
\end{figure}

In the first part of this section we use the comparison between the two motors diffusion coefficients to measure the breakdown of the scallop theorem.
Then in the second part of the section we use the efficiency of motion for a more precise investigation of the  breakdown domains.
Figure \ref{f3} shows that for small  force field moments, the diffusion coefficient {\color{black}$D$ and the coefficient $\kappa$ are} very small for the time symmetric motor and roughly constant for the time asymmetric motor. As discussed in the previous section for these force fields the scallop theorem is not broken for the symmetric motor and the diffusion is roughly independent on the field for the asymmetric motor. 
We then observe a  moment threshold around $100 pN.$\AA\ above which the diffusion increases.
 Figure \ref{ff00} shows that the threshold is of the order of magnitude of the Brownian noise on the moment.
As a result, the threshold corresponds to the {\color{black} external force} value that counterbalance the Brownian forces fluctuations at the temperature of study.  
Around the threshold the diffusion is the same for both motors. Because the difference between the two motors is related to the Purcell theorem, the equivalence of the diffusion coefficients shows that the field induces a breakdown of the Purcell theorem for the time symmetric motor.
The field evolution of the symmetric motor's efficiency in Figure \ref{f5} confirms that picture.
These results suggest that when the external forces are  comparable to the Brownian forces on the motor, it induces a fluctuating asymmetry that breaks the theorem.

Then around $500 pN.$\AA\ the diffusion  stabilizes to $D\approx1$\AA$^{2}/ns$ for the time symmetric motor and  $D\approx3$\AA$^{2}/ns$ for the time asymmetric motor. 
That difference between the two saturation values suggests a decrease of the efficiency of the symmetric motor for large fields, in agreement with the results displayed in Figure \ref{f5}.
To resume that picture, for small force fields the sum of Brownian fluctuations and of the field leads to an asymmetry of motion that induces the breakdown of the Purcell theorem, while for large fields, the field finally induces a one dimension motion, leading to a saturation and the observed flat curves.

We verify this picture in Figure \ref{f5} that shows the motion's efficiency $\epsilon$ (a quantity that quantifies the breakdown of Purcell's theorem)\cite{prefold} versus {\color{black} the force field moment} for the time symmetric motor.
The efficiency of the time symmetric motor in figure \ref{f5} first increases at the threshold value ({\color{black}for a moment $M\approx 100pN.$\AA}) and then decreases for large fields to its  zero field value.
Therefore the scallop theorem holds approximately ($\epsilon=0.3$) for small fields then is   broken above a threshold field value and finally is restored for large fields.
 As a result, {\color{black}as observed in Figure \ref{f5}},  the efficiency is  maximum when the force field is of the same order of magnitude than the Brownian force fluctuations, while large and small fields do not modify the efficiency.
 To conclude, for small fields, the effect of the field acts as a small perturbation inside the Brownian motion while 
 large fields destroy the fluctuations restoring the scallop theorem for the symmetric motor. Only for fields of the same order of magnitude  than the Brownian fluctuations, the field perturbation induces a significant breakdown of the scallop theorem. 
{\color{black} However for large force fields the motor's diffusion  nonetheless increases in Figure \ref{f3} due to the orientation of the motor's motion.}

\begin{figure}
\centering
\includegraphics[height=6.6 cm]{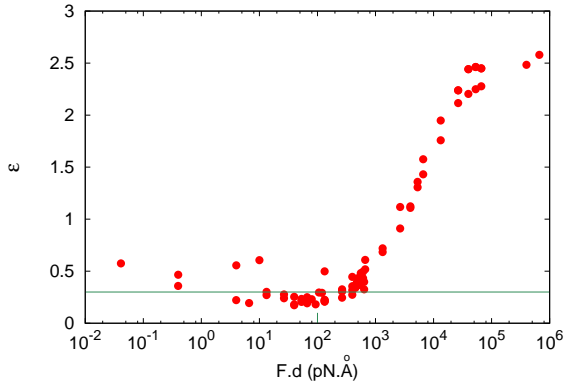}
\caption{(color online) Mobility efficiency coefficient $\displaystyle{\epsilon= {<r^{2}(n\tau_{p})>\over 2n<r^{2}(\tau_{p}/2)>} }$ with $n=3$, versus the maximum moment for the time asymmetric motor.
Values larger than $1$ imply that the motor has a preferential direction.} 
\label{f6}
\end{figure}

The time asymmetric motor's efficiency behaves differently (see Figure 6).
It increases at the same threshold value, but do not decreases for large fields saturating instead to $\epsilon \approx 2.6$.
As the breakdown of the scallop theorem for the time asymmetric motor doesn't relies only on Brownian motion but on the time asymmetry of the motor,
the scallop theorem is not restored for large fields, explaining the large efficiency for large fields.
{\color{black} The large values are here due to the orientation of the motor towards the field.}
Eventually, this difference between the motor's efficiencies explains the difference between the motor's diffusion coefficients observed in Figure \ref{f3}.

\begin{figure}
\centering
\includegraphics[height=6.6 cm]{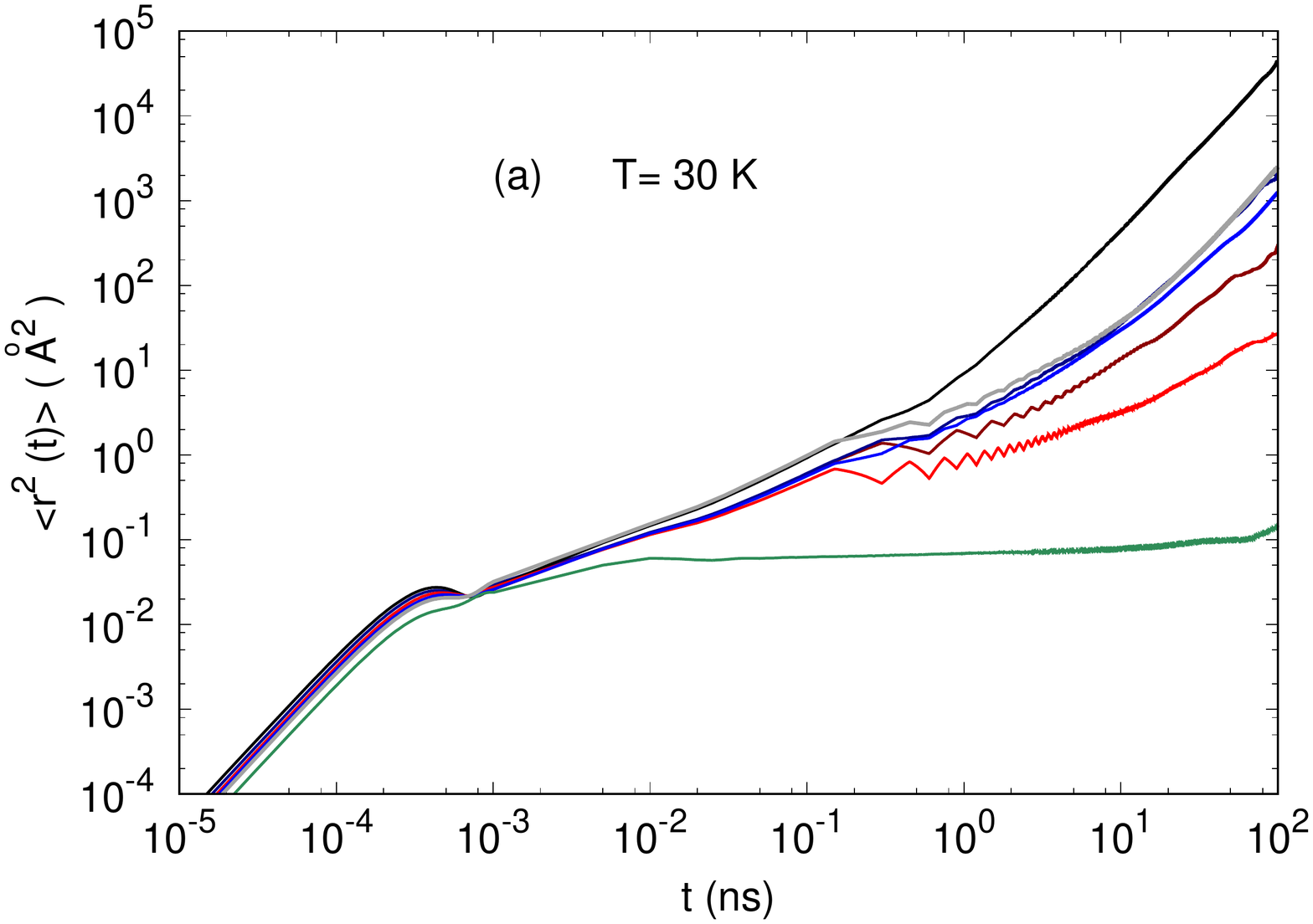}
\includegraphics[height=6.6 cm]{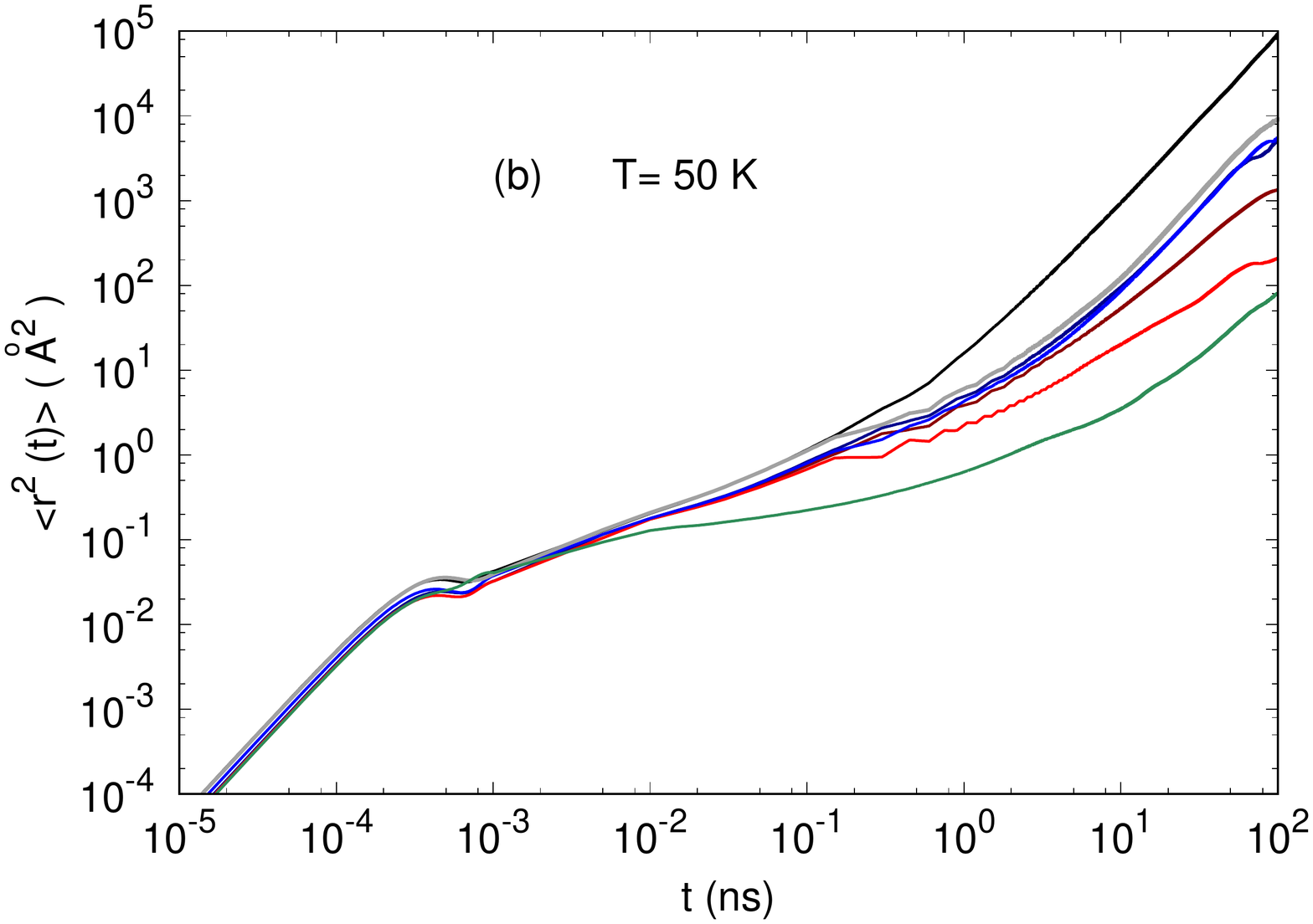}
{\color{black} \caption{(color online) Motor's mean square displacement $<r^{2}(t)>$ (a)  At a temperature for which the medium without stimuli is solid  T=30K, (b) the medium is a viscous liquid T=50K.
In each Figure, from bottom to top the curves correspond to: The motor off and no field (dark green curve); the motor on: no field ($F.d$=0), time symmetric motor (red curve), then asymmetric (dark red curve);   $F.d=6.6$ $10^{2} pN.$\AA\ : symmetric (blue curve), then asymmetric (dark blue curve) these two curves superimpose almost perfectly for both temperatures; $F.d= 6.6$ $10^{3} pN.$\AA\ time symmetric (gray curve) then asymmetric (black curve).
}
\label{f3x}
}
\end{figure}

{\color{black}
\subsection{Comparison of the two motor's displacements}

In Figure \ref{f3x} we show the two motor's mean square displacements (MSD) for different force fields and temperatures.
Figure \ref{f3x}a corresponds to a temperature for which the medium is solid when the motor is off.
The dark-green curve (bottom) is flat showing that the medium is actually solid when the motor is off.
At very short time scales, the ballistic regime gives us the average temperature of the motor.
Figure \ref{f3x}a shows that the motor's temperature is $15\%$ larger when the motor is active  than inactive while we didn't find any difference for the medium around it.
However  the ballistic part of the curves (giving the temperature) was calculated during the foldings only and do not take into account the long relaxation.
Therefore the motor's temperature is actually smaller than the ballistic regime suggests in the Figure.

The Figure shows that the time asymmetric motor is the more efficient of the two motors, for weak and large fields, but not for fields in between. The blue and dark blue curves superimpose, showing that for the corresponding field range the time asymmetric and symmetric motors are equivalent.
We observe the same results at a larger temperature in Figure \ref{f3x}b when the medium is liquid. These results are in agreement with the time symmetric motor's  maximum efficiency for intermediate fields observed in Figure \ref{f5}, and with the comparison between diffusive coefficients in Figure \ref{f3}.

When the external field gives a privileged direction to the motor, we observe displacements larger than diffusive (see for example the end of the black curve on the top) as expected for a directional motion. 
For the time symmetric motor, we expect these super-diffusive behaviors to disappear for larger time scales, due to symmetry (as the $X$ and $-X$ directions are equivalent for that motor).
The super-diffusive behavior demonstrates that the motor moves by its own, as a motor carried by the medium would be purely diffusive.
}

\subsection{Elementary displacements of the nano-swimmer}

\begin{figure}
\centering
\includegraphics[height=6.6 cm]{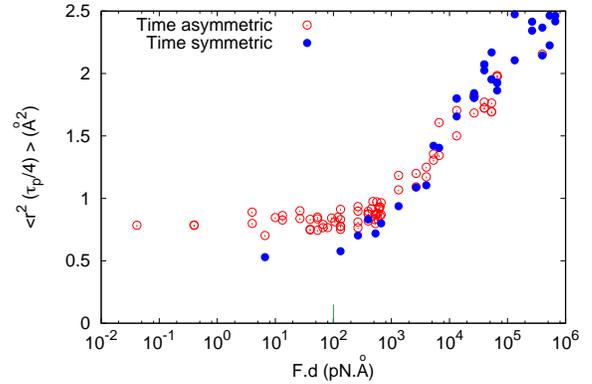}
\caption{(color online) Mean square displacement of the motors after one flap ($\Delta t=\tau_{p}/4$) versus the maximum moment.} 
\label{f7}
\end{figure}

\begin{figure}
\centering
\includegraphics[height=6.6 cm]{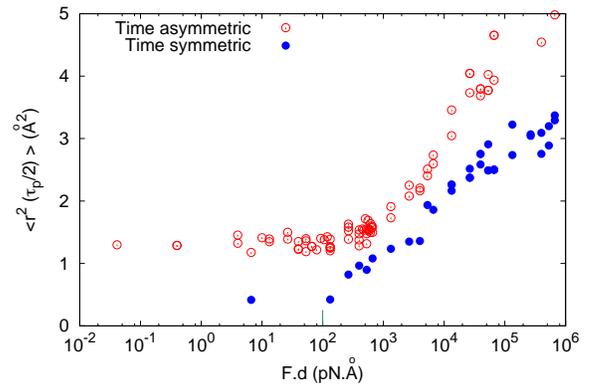}
\caption{(color online) Mean square displacement of the motors after the second flap ($\Delta t=\tau_{p}/2$) versus the maximum moment.} 
\label{f8}
\end{figure}
\begin{figure}
\centering
\includegraphics[height=6.6 cm]{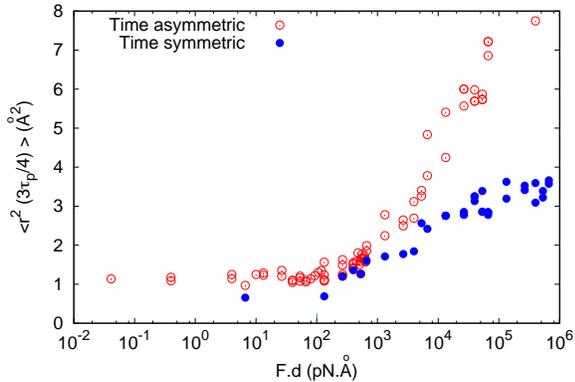}
\caption{(color online) Mean square displacement of the motors after the third flap ($\Delta t=3\tau_{p}/4$) versus the maximum moment.} 
\label{f9}
\end{figure}
\begin{figure}
\centering
\includegraphics[height=6.6 cm]{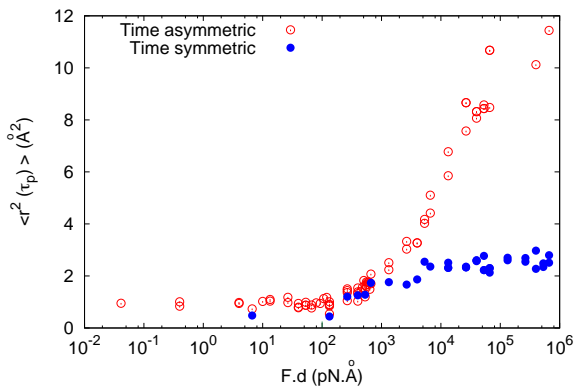}
\caption{(color online) Mean square displacement of the motors after the last flap ($\Delta t=\tau_{p}$) versus the maximum moment.} 
\label{f10}
\end{figure}

In this section we describe the motors displacements following each arm flap.
The elementary displacements after the first to the fourth flaps of the motors {\color{black} are displayed respectively} in Figures \ref{f7}, \ref{f8}, \ref{f9} and \ref{f10}.
After the first motor's flap in Figure \ref{f7}, the time symmetric and time asymmetric motors behave the same way as the time asymmetry didn't still act.
$<r^{2}(\tau_{p}/4)>$ is small and constant below the force field moment threshold and then {\color{black} increases continuously above the threshold.}

After the second flap, {\color{black} we observe a small difference between the two motors average displacements} in Figure \ref{f8}.
{\color{black} Notice that due to the averaging on the time origin, the second flap is not similar in the displacement calculations for the two motors.}
While for the time symmetric motor the displacements are only slightly larger than after the first flap, for  the time asymmetric motor the displacements are larger even at zero field.
{\color{black}The difference between the displacements of the two motors is approximately constant on the Figure.}

After the third flap in Figure \ref{f9}, the displacement of the time symmetric motor increases and becomes approximately equal to the time asymmetric motor displacement around the threshold.
For large fields in contrast the time asymmetric motor displacement increases rapidly  to values much larger than the time symmetric motor's displacements.
{\color{black} The alignment of the motor towards the field leads to much larger displacements for the time asymmetric motor, as the scallop theorem begins to apply for that third step on the symmetric motor leading to backward motions.}

Eventually after the fourth and last flap in Figure \ref{f10}, the mean square displacements of the two motors is similar  up to the threshold field's value and separate for large fields, the time asymmetric motor displacement being much larger.
Notice that the time symmetric motor's displacement for large field after the last flap is approximately equal to its displacement after the first flap ($2.5$\AA$^{2}$), showing that  the scallop theorem applies on the time symmetric motor.

Notice that Brownian noise forces are rather large in our conditions, leading to a large external force field to counterbalance it.
The Brownian noise forces can be lowered using a medium at lower pressure, but it will also increase at higher temperature.
The threshold external torque in our system is around $100pN.$\AA. It corresponds {\color{black} in an example motor molecule (A typical azobenzene molecule is approximately $15$ \AA\ long)} to opposite forces of $10pN$ separated by a $10$ \AA\ distance.
{\color{black} We used a shorter distance $d=4$\AA\ in our simulations in order to put the forces on non moving parts of the molecule (to insure the observed effects are not generated by the work induced by external forces on motor's moving parts). }
For an electric charge $q=e${\color{black}, $10$pN forces} correspond to an electric field of $6. 10^{7} V/m$.
Fields of that magnitude or larger can be encountered near polarized interfaces for example.
{\color{black} The motor's alignment can also be generated by the environment if the long motor is embedded inside a nematic or smectic liquid crystal
 (see for example ref.\cite{lc1,lc2,lc3}), the interactions with the liquid crystal molecules orienting the motor.
 Eventually, the alignment can also be generated by confinement inside nanopores or on surface walls\cite{conf1}.
}

 For application purposes our motor can be created with two chemically bounded azobenzene molecules or derivatives.
 That flat molecule in its trans conformation has the property of photo-isomerization, folding into its cis conformation when subjected to a light stimulus.
  The molecule then relaxes to its trans conformation, a relaxation that can be accelerated by a second light stimulus. For further information on azobenzene photo-isomerization properties and applications, see the sound authoritative review by Natansohn and Rochon ref.\cite{azo1} {\color{black} and recent applications \cite{az1,lc1,flu1,flu2,flu3,conf1}. 
}
\vskip 1cm
\section{Conclusion}

In this work we have used molecular dynamics simulations to study the effect of an electric field on the motion of a polarized molecular motor (nano-swimmer).
Our objectives were to induce a breakdown of the scallop theorem from the anisotropic motion induced by the electric field and to orient the motion of the swimmer in the direction of the field.
To evaluate the extent of the scallop theorem  {\color{black} breakdown}, we compared the motions of a nano-swimmer with a time reversible sequence of flaps, with the motion of the same swimmer with a non-reversible sequence of flaps. 

We found a field threshold for the orientation of the swimmer's motion that is the same for both swimmers.
The two swimmers share the same displacements around the threshold, showing that the Purcell's theorem is also broken for the time symmetric swimmer in that field range.
In agreement with that interpretation, the mobility efficiency $\epsilon$ measuring the extent of the breakdown of the theorem displays a wide peak around that field range for the time symmetric swimmer.
Therefore there is a field range for which, due to the competition of the orientation field and Brownian forces, Purcell's theorem is broken for the time symmetric motor. 
For larger fields, the motion of the swimmer is oriented in the direction of the field, increasing the displacements even for the time symmetrical swimmer hindered by Purcell's theorem.
Then the swimmer's orientation saturates as the swimmer tends to be totally oriented in the direction of the field.
For small and large fields, the time asymmetrical swimmer is more efficient, as expected by Purcell\cite{scallop1}.
Eventually, our results suggest a development in nano-swimmers embedded inside liquid crystals with and without electric fields.


\begin{thebibliography}{99}





\bibitem{moto1} J.R. Howse, R.A.L. Jones, A.J. Ryan, T. Gough, R. Vafabakhsh, R. Golestanian, 
 \newblock  {Phys. Rev. Lett. }  {\bf 99}, 048102 (2007).

\bibitem{moto2} H. Karani, G.E. Pradillo, P.M. Vlahovska, 
 \newblock  {Phys. Rev. Lett. }  {\bf 123}, 208002 (2019).

\bibitem{moto3} F. Novotny, M. Pumera, 
 \newblock  {Scien. Rep. }  {\bf 9}, 13222 (2019).

\bibitem{moto4} X. Arque, A. Romero-Rivera, F. Feixas, T. Patino, S. Osuna, S. Sanchez,
 \newblock  {Nature Comm. }  {\bf 10}, 2826 (2019).

\bibitem{moto5} A.M. Brooks, M. Tasinkevych, S. Sabrina, D. Velegol, A. Sen, K.J.M. Bishop, 
 \newblock  {Nature Comm. }  {\bf 10}, 495 (2019).

\bibitem{moto6} P. Pietzonka, E. Fodor, C. Lohrmann, M.E. Cates, U. Seifert, 
 \newblock  {Phys. Rev. X}  {\bf 9}, 041032 (2019).

\bibitem{moto7} C. Calero, J. Garcia-Torres, A. Ortiz-Ambriz, F. Sagues, I. Pagonabarraga, P. Tierno,
 \newblock  {Nanoscale}  {\bf 11}, 18723 (2019).



\bibitem{motor0} H. Hess, 
 \newblock  {Annu. Rev. Biomed. Eng. }  {\bf 13}, 429-450 (2011).


\bibitem{motor1} R.D. Astumian, 
 \newblock  {Science }  {\bf 276}, 917-922 (1997).

\bibitem{motor2} M.F. Hawthorne, J.I. Zink, J.M. Skelton, M.J. Bayer, C. Liu, E. Livshits, R. Baer, D. Neuhauser, 
 \newblock  {Science }  {\bf 303}, 1849-1851 (2004).

\bibitem{motor3} P. Palffy-Muhoray, T. Kosa, E. Weinan, 
 \newblock  {Appl. Phys. A} {\bf 75}, 293-300 (2002).

\bibitem{motor4} J. Berna, D.A. Leigh, M. Lubomska, S.M. Mendoza, E.M. Perez, P. Rudolf, G. Teobaldi, F. Zerbetto, 
 \newblock  {Nature Mater.}  {\bf 4}, 704-710 (2005).

\bibitem{motor5} T.R. Kline, W.F. Paxton, T.E. Mallouk, S. Ayusman,
 \newblock  {Angew. Chem. Int. Ed.}  {\bf 44}, 744-746 (2005).

\bibitem{motor6} W.R. Browne, B.L. Feringa, 
 \newblock  {Nature Nanotech.}  {\bf 1}, 25-35 (2006).

\bibitem{motor7} K. Dholakia, P. Reece, 
 \newblock  {Nanotoday}  {\bf 1}, 20-27 (2006).


\bibitem{motor8} T. Fehrentz, M. Schonberger, D. Trauner, 
 \newblock  {Angew. Chem. Int. Ed.} {\bf 50}, 12156-12182 (2011).


\bibitem{motor10} M.M. Russew, S. Hecht,
 \newblock  {Adv. Mater.} {\bf 22}, 3348-3360 (2010).

\bibitem{motor11} N. Katsonis, M. Lubomska, M.M. Pollard, B.L. Fearing, P. Rudolf,
 \newblock  {Progress in Surface Science}  {\bf 82}, 407-434 (2007).


\bibitem{motor12} A.P. Davis,
 \newblock  {Nature}  {\bf 401}, 120-121 (1999).

\bibitem{motor13} J.P. Sauvage,
 \newblock \emph{Molecular machines and motors} , {Springer, Berlin},{ 2001}.


\bibitem{motor14} E.R. Kay, D.A. Leigh,
 \newblock  {Nature}  {\bf 440}, 286-287 (2006).


\bibitem{motor15} V. Balzani, et al.,
 \newblock  {Proc. Nat. Acad. Sci.}  {\bf 103}, 1178-1183 (2006).


\bibitem{motor16} T. Muraoka, K. Kinbara, Y. Kobayashi, T. Aida,
 \newblock  {JACS}  {\bf 125}, 5612-5613 (2003).


\bibitem{motor17} T.J. Huang, et al.,
 \newblock  {Nano Lett.}  {\bf 4}, 2065-2071 (2004).

{\color{black}
 \bibitem{pccp}  S. Ciobotarescu, N. Hurduc, V. Teboul,
\newblock  {  Phys. Chem. Chem. Phys.}  {\bf  18}, 14654 (2016).}

\bibitem{motor18} A.S. Amrutha, K.R.S. Kumar, T. Kikukawa, N. Tamaoki,
 \newblock  {ACS Nano}  {\bf 11}, 12292-122301 (2017).




\bibitem{scallop1} E. M. Purcell,
 \newblock  {Am. J. Phys.} {\bf 45}, 3 (1977)


\bibitem{purc}
{\color{black} Purcell original paper demonstrates the 'scallop theorem' for a continuous medium subjected to Stokes equations. However the theorem holds in a much larger domain. Therefore we use the term 'breakdown' (of the theorem) when special conditions induce a departure from its predictions, if the theorem holds without these special conditions even if we are not in the original conditions of applications of Purcell's paper.}


{\color{black}
\bibitem{prefold}  S. Ciobotarescu, S. Bechelli, G. Rajonson, S. Migirditch, B. Hester, N. Hurduc, V. Teboul,
 \newblock {Phys. Rev. E}   {\bf 96}, 062614 (2017)

\bibitem{scallop13b} V. Teboul, G. Rajonson,
 \newblock {Phys. Chem. Chem. Phys.}  {\bf 21}, 2472 (2019)
 
 \bibitem{scallop13c} V. Teboul, G. Rajonson,
 \newblock {J. Chem. Phys.}  {\bf 150}, 144502 (2019)

 
 }

\bibitem{scallop2}  T. Qiu, et al.,
 \newblock  {Nature Comm.} {\bf 5}, 5119 (2014)




\bibitem{scallop4} E. Lauga,
 \newblock {Physics of Fluids}  {\bf 19}, 061703 (2007)

\bibitem{scallop7} E. Lauga,
 \newblock {Soft Matter}  {\bf 7}, 3060 (2011)


\bibitem{scallop5} D. Du, E. Hilou, S. L. Biswal,
 \newblock {Soft Matter}  {\bf 14}, 3463 (2018)

\bibitem{scallop6} S. Alben, M. Shelley,
 \newblock {PNAS}  {\bf 102}, 11163 (2005)

\bibitem{scallop8} M. Theers, R. G. Winkler,
 \newblock {Soft Matter}  {\bf 10}, 5894 (2014)

\bibitem{scallop9} P. Olla,
 \newblock {Eur. Phys. J. B}  {\bf 80}, 263 (2011)

 
\bibitem{scallop11} P. Olla,
 \newblock {Phys. Rev. E}  {\bf 89}, 032136 (2014)
 

\bibitem{scallop12} E. Lauga,  D. Bartolo,
 \newblock {Phys. Rev. E}  {\bf 78}, 030901R (2008)

\bibitem{scallop13} M.F. Lapa, T.L. Hugues,
 \newblock {Phys. Rev. E}  {\bf 89}, 043019 (2014)
 
 
 \bibitem{scallop14} S. Childress, R. Dudley,
 \newblock {J. Fluid. Mech.}  {\bf 498}, 257 (2004)

\bibitem{scallop15} X. Lu, Q. Liao,
 \newblock {Phys. Fluids}  {\bf 18}, 098104 (2006)


\bibitem{scallop10} P. Olla,
 \newblock {Phys. Rev. E} {\bf 82}, 015302R (2010)


 \bibitem{scallop3} E. Lauga,
\newblock  {Phys.Rev.Lett.} {\bf 106}, 178101 (2011)

{\color{black}
\bibitem{md16}  V. Teboul, M. Saiddine, J.M. Nunzi, 
\newblock  { Phys. Rev. Lett. }  {\bf  103}, 265701 (2009)


\bibitem{carry}  M. Saiddine, V. Teboul, J.M. Nunzi, 
 \newblock {J. Chem. Phys. }   {\bf133}, 044902 (2010)


\bibitem{cage}  V. Teboul, M.  Saiddine, J.M. Nunzi, J.B. Accary, 
 \newblock {J. Chem. Phys. }  {\bf134}, 114517 (2011)


\bibitem{rate}  J.B. Accary, V. Teboul,   
{J. Chem. Phys.}  {\bf139}, 034501 (2013)
}

\bibitem{flu1} P. Karageorgiev, D. Neher, B. Schulz, B. Stiller, U. Pietsch, M. Giersig, L. Brehmer, 
 {Nature Mater.}  {\bf4}, 699-703 (2005).

\bibitem{flu2} G.J. Fang, J.E. Maclennan, Y. Yi, M.A. Glaser, M. Farrow, E. Korblova, D.M. Walba, T.E. Furtak, N.A. Clark, 
 {Nature Comm.}  {\bf4}, 1521 (2013).

\bibitem{flu3} N. Hurduc, B.C. Donose, A. Macovei, C. Paius, C. Ibanescu, D. Scutaru, M. Hamel, N. Branza-Nichita, L. Rocha,
 {Soft Mat.}  {\bf 10}, 4640-4647 (2014).

\bibitem{flu4}  J. Vapaavuori, A. Laventure, C.G. Bazuin, O. Lebel, C. Pellerin, 
 { J. Amer. Chem. Soc.}  {\bf 137}, 13510 (2015).

\bibitem{gt1}  K. Binder, W. Kob,
 {\em Glassy materials and disordered solids}, World Scientific, Singapore 2011.

\bibitem{gt2}  L. Berthier, G. Biroli, J.P. Bouchaud, L. Cipelletti, W. Van Saarloos,
 {\em Dynamical heterogeneities in glasses, colloids and granular media}, Oxford Science Publications, Oxford 2011.


{\color{black}
\bibitem{swim} J.E. Avron, O. Raz, 
 \newblock  {New Journal of Physics } {\bf10}, 063016 (2008).}


\bibitem{flu5}  V. Teboul, R. Barille, P. Tajalli, S. Ahmadi-Kandjani, H. Tajalli, S. Zielinska,  E. Ortyl,
\newblock  {  Soft Matt.}  { \bf   11}, 6444 (2015).



\bibitem{azo1} A. Natansohn,  P. Rochon, 
 \newblock  {Chem. Rev. } {\bf102}, 4139-4175 (2002).
 

\bibitem{azo2} J.A. Delaire, K. Nakatani, 
 \newblock  { Chem. Rev.} {\bf 100}, 1817 (2000).
 
\bibitem{azo2b} G.S. Kumar, D.C. Neckers,  
\newblock  { Chem. Rev.} {\bf 89}, 1915 (1989).

\bibitem{azo3} K.G. Yager, C.J. Barrett,  
\newblock  {Curr. Opin. Solid State Mater. Sci.} {\bf 5}, 487 (2001).



\bibitem{azo4}  T.G. Pedersen, P.M.  Johansen, 
\newblock  {  Phys. Rev. Lett.}  {\bf  79}, 2470-2473 (1997).

\bibitem{azo5}  T.G. Pedersen, P.M.  Johansen, N.C.R.  Holme, P.S.  Ramanujam, S. Hvilsted,
\newblock  {  Phys. Rev. Lett.}  {\bf  80}, 89-92 (1998).

\bibitem{azo6}  J. Kumar, L.  Li, X.L.  Jiang, D.Y.  Kim, T.S.  Lee, S. Tripathy, 
\newblock  {  Appl. Phys. Lett.}  {\bf  72}, 2096-2098 (1998).



\bibitem{azo7}  C.J. Barrett, P.L.  Rochon, A.L.  Natansohn, 
\newblock  {  J. Chem. Phys.}  {\bf  109}, 1505-1516 (1998).

\bibitem{azo8}   C.J. Barrett, A.L. Natansohn, P.L. Rochon, 
\newblock  {  J. Phys. Chem.} {\bf  100}, 8836-8842 (1996).

\bibitem{azo9}  P. Lefin, C.  Fiorini,  J.M. Nunzi, 
\newblock  {  Pure Appl. Opt.}  {\bf  7}, 71-82 (1998).



\bibitem{md1} M.P. Allen,   D.J. Tildesley, 
\emph {Computer Simulation of Liquids}, Oxford University Press, New York 1990.

\bibitem{md2} M. Griebel, S. Knapek, G. Zumbusch, 
\emph {Numerical Simulation in Molecular Dynamics}, Springer-Verlag, Berlin 2007.

\bibitem{md2b} D. Frenkel, B. Smit, 
\emph {Understanding Molecular Simulation}, Academic Press, San Diego 1996.

{\color{black}
\bibitem{md4}  
S. Chaussedent, V. Teboul, A. Monteil,
%
\newblock {Curr. Opin. Solid State Mater. Science}  {\bf 7},  111-116 (2003)
}


\bibitem{ms1}
F. Ritort, P. Sollich, 
Adv. Phys.  {\bf 52}, 219 (2003)

\bibitem{ms2}
G. H. Fredrickson, H. C. Andersen, 
Phys. Rev. Lett. {\bf 53}, 1244 (1984)

\bibitem{ms3}
J. Jackle,  S. Eisinger, 
Z. Phys. B  {\bf 84}, 115 (1991)

\bibitem{ms4}
W. Kob and H. C. Andersen, 
Phys. Rev. E  {\bf 48}, 4364 (1993)

{\color{black}
\bibitem{ms5}  
V. Teboul,
\newblock {J. Chem. Phys.}  {\bf 141},  194501 (2014)}


\bibitem{md3}  
C.F.E. Schroer, A. Heuer, 
   {Phys. Rev. Lett.} {\bf  110}, 067801 (2013).


{\color{black}
\bibitem{md4b}  
J.B. Accary, V. Teboul, 
\newblock  {  J. Chem. Phys.} {\bf  136}, 0194502 (2012).}




\bibitem{md6}  
A. Furukawa, K. Kim, S. Saito, H. Tanaka, 
  { Phys. Rev. Lett.} {\bf  102}, 016001 (2009).

\bibitem{md7}  
T. Iwashita, T. Egami, 
  { Phys. Rev. Lett.} {\bf  108}, 196001 (2012).

\bibitem{md8}  
T. Gleim, W. Kob, K. Binder, 
  { Phys. Rev. Lett.} {\bf  81}, 4404 (1998).

\bibitem{md9}  
E.J. Saltzman, K.S. Schweitzer, 
  {J. Chem. Phys.} {\bf  125}, 044509 (2006).


{\color{black}
\bibitem{md10}  V. Teboul, Y. Le Duff,
 \newblock  {   J. Chem. Phys.}  {\bf 107}, 10415-10419 (1997).


\bibitem{yld} Y. Le Duff, V. Teboul, 
%
 \newblock  { Phys. Lett. A}  {\bf 157}, 44 (1991).
}



\bibitem{md11}  
B. Uralcan, I.A. Aksay, P.G. Debenedetti, D. Limmer,
\newblock  {  J. Phys. Chem. Lett.}  { \bf  7}, 2733 (2016).

\bibitem{md12}  
J. Zhang, F. Muller-Plathe, F. Leroy,
\newblock  {  Langmuir}  { \bf  31}, 7544 (2015).



\bibitem{md13}  
E. Flenner, G. Szamel,
\newblock  {  J. Phys. Chem. B}  { \bf  119}, 9188 (2015).

\bibitem{md14}  
E. Flenner, G. Szamel,
\newblock  {  Nature Comm.}  { \bf  6}, 7392 (2015).

\bibitem{md15}  
Y.S. Elmatad, A.S. Keys,
\newblock  {  Phys. Rev. E}  { \bf  85}, 061502 (2012).

{\color{black}
\bibitem{c2}  V. Teboul, 
\newblock  {  Eur. Phys. J. B}  { \bf   51}, 111 (2006).}


\bibitem{keys}  
A. S. Keys, L. O. Hedges, J.P. Garrahan, S.C. Glotzer, D. Chandler,  
\newblock  {  Phys. Rev. X}  { \bf   1}, 021013 (2011).


{\color{black}
\bibitem{ariane} A.P. Kerasidou, Y. Mauboussin, V. Teboul, 
 {Chem. Phys.} {\bf 450}, 91 (2015).}

 
\bibitem{berendsen}  H.J.C. Berendsen,  J.P.M. Postma, W. Van Gunsteren,  A. DiNola, J.R.  Haak, 
 \newblock  {  J. Chem. Phys.}  {\bf 81}, 3684-3690 (1984).

{\color{black}
\bibitem{finite2}  
S. Taamalli, J. Hinds, S. Migirditch, V. Teboul,
\newblock {Phys. Rev. E}  {\bf 94},  052604 (2016)}

 
\bibitem{mix1}  R.J. Good, C.J. Hope,
 \newblock {J. Chem. Phys.}  {\bf53}, 540 (1970).

\bibitem{mix2}  J. Delhommelle, P. Millie,
\newblock {Mol. Phys.}  {\bf99}, 619 (2001).




\bibitem{fragile1}  C.A. Angell, 
\newblock  {  Science}  { \bf 267}, 1924-1935 (1995).


\bibitem{fragile2}  N.A. Mauro, M. Blodgett, M.L. Johnson, A.J. Vogt, K.F. Kelton, 
\newblock  { Nature Comm.}  { \bf 5}, 4616 (2014).

 
{\color{black}
\bibitem{lc1} M. Hada, D. Yamaguchi, T. Ishikawa, T. Sawa, K. Tsuruta, K. Ishikawa, S. Koshihara, Y. Hayashi, T. Kato, 
\newblock  { Nature Comm.}  { \bf 10}, 4159 (2019).



\bibitem{lc2} R. Rosenbauer, et al., 
\newblock  { Macromolecules}  { \bf 38}, 2213 (2005).



\bibitem{lc3} T. Ikeda, O. Tsutsumi, 
\newblock  {Science}  { \bf 268}, 1873 (1995).


\bibitem{conf1} M. Muller, J. Henzl, K. Morgenstern,
\newblock  {Chem. Phys. Lett.}  { \bf 738}, 136906 (2020).

\bibitem{az1} Z. Chu, R. Klajn,
\newblock  {Nano Lett.}  { \bf 19}, 7106 (2019).

}







\end{thebibliography}
\end{document}